%% file: eprint.tex
\documentclass[10pt]{article}
\usepackage{graphicx}

\usepackage{verbatim}   
\usepackage{color}      
\usepackage{subfigure}  
\usepackage[abs]{overpic}
\usepackage{float}
\restylefloat{table}
\usepackage{lineno}     

\def\Title#1{\begin{center} {\Large #1 } \end{center}}
\def\Author#1{\begin{center}{ \sc #1} \end{center}}
\def\Address#1{\begin{center}{ \it #1} \end{center}}

\newcommand\pubblock{\rightline{\begin{tabular}{l} Proceedings of the Fifth Annual LHCP\\ \pubnumber\\
         \pubdate  \end{tabular}}}

\newenvironment{Abstract}{\begin{quotation} \begin{center} 
             \large ABSTRACT \end{center}\bigskip 
      \begin{center}\begin{large}}{\end{large}\end{center} \end{quotation}}

\newenvironment{Presented}{\begin{quotation} \begin{center} 
             PRESENTED AT\end{center}\bigskip 
      \begin{center}\begin{large}}{\end{large}\end{center} \end{quotation}}

\input econfmacros.tex

\usepackage{color}

\newcommand\pubnumber{ ATL-PHYS-PROC-2017-131 }

\newcommand\pubdate{\today}

\def\affiliation{
On behalf of the ATLAS Collaboration, \\
INFN Gruppo Collegato di Udine, Sezione di Trieste, Udine and ICTP, Trieste \\
Strada Costiera 11, Trieste 34151, Italy}

\begin{document}

\large
\begin{titlepage}
\pubblock

\vfill
\Title{Search for the SM Higgs boson produced in association with top quarks at $\sqrt{s}$ = 13 TeV with the ATLAS detector at the LHC}
\vfill

\Author{ Leonid Serkin }
\Address{\affiliation}
\vfill
\begin{Abstract}
A review of the searches for the SM Higgs boson produced in association with a pair of top quarks, \ttH, using up to 13.3~fb$^{-1}$ of $pp$ collisions at $\sqrt{s}$ = 13 TeV collected with the ATLAS detector at the LHC is presented. Searches in the diphoton, multilepton and \bbbar\ decay channels are summarised. The combination of these searches yields a measured (expected) significance for the observation of the \ttH\ production process of $2.8 \sigma$ ($1.8 \sigma$).

\end{Abstract}
\vfill

\begin{Presented}
The Fifth Annual Conference\\
 on Large Hadron Collider Physics \\
Shanghai Jiao Tong University, Shanghai, China\\ 
May 15-20, 2017
\end{Presented}
\vfill
\end{titlepage}
\def\thefootnote{\fnsymbol{footnote}}
\setcounter{footnote}{0}
%

\normalsize

\section{Introduction}
Due to its small production cross-section ($\sim 1\%$ of the total Higgs boson cross-section), the SM Higgs boson production in association with a pair of top quarks, \ttH, has not yet been observed. The measurement of the \ttH\ production rate would provide a direct measurement of the Yukawa coupling of the top quark to the Higgs boson and is instrumental in determining ratios of Higgs boson couplings in a model independent way. 
The \ttH\ production can be studied in a variety of final state topologies, depending on the top quark decay topology and the Higgs boson decay mode. In the following, we review the searches for the \ttH\ production in the diphoton (\ttHgg), multilepton (\ttHleptons) and \ttHbb\ decay channels, performed using up to $13.3$~fb$^{-1}$ of $pp$ collisions at $\sqrt{s}$ = 13 TeV collected with the ATLAS detector~\cite{bib:atlas} at the LHC.

\section{\ttHgg\ analysis overview}
The Higgs boson decay into two photons ($H \to \gamma \gamma$) is a particularly attractive way to study the properties of the Higgs boson. Despite the small branching ratio, a reasonably large signal yield can be obtained thanks to the high photon reconstruction and identification efficiency in ATLAS. Furthermore, due to the excellent photon energy resolution of the ATLAS calorimeter, the signal manifests itself as a narrow peak in the diphoton invariant mass spectrum on top of a smoothly falling background, and the Higgs boson signal yield can be measured using an appropriate fit.

\begin{figure*}[h]
\centering
\begin{overpic}[height=0.27\textwidth]{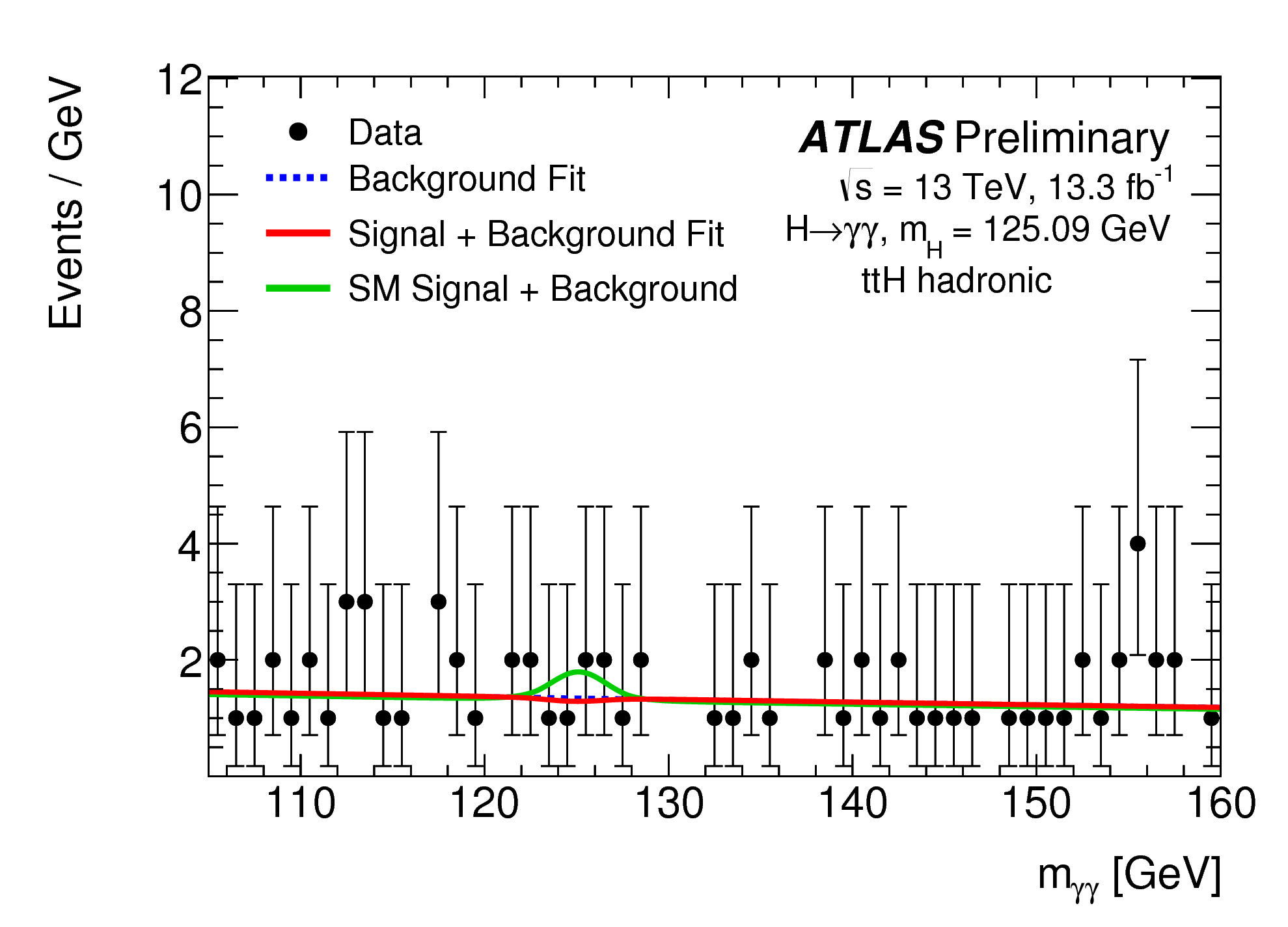}
\put(95,-7){(a)}
\end{overpic}
\qquad
\qquad
\begin{overpic}[height=0.27\textwidth]{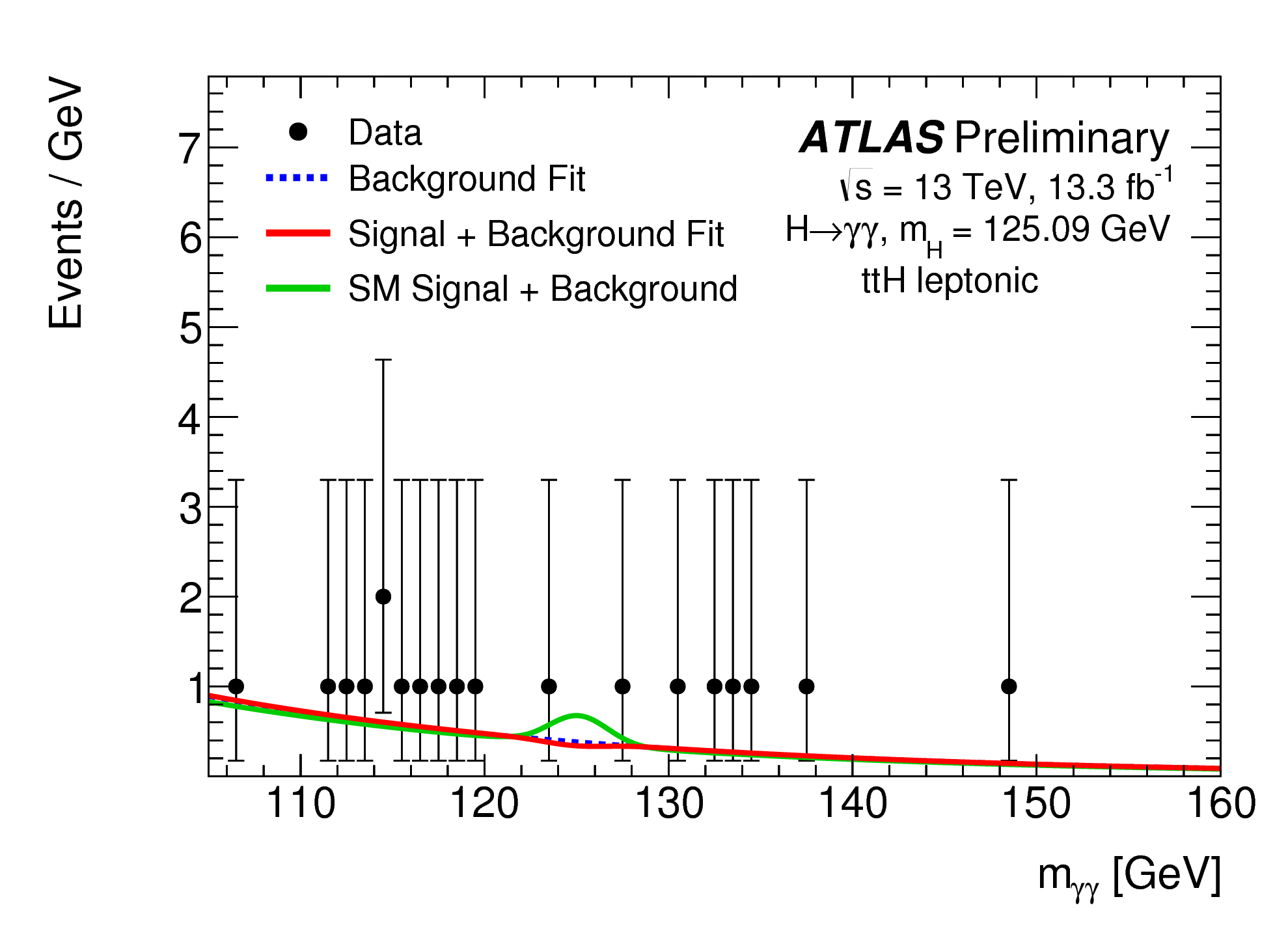}
\put(95,-7){(b)}
\end{overpic}
\caption{Diphoton invariant mass spectrum in the \ttHgg\ search: (a) hadronic and (b) leptonic top decay modes~\cite{bib:ttH_gg}.}
\label{fig:Plots1} 
\end{figure*}

The events selected in the diphoton baseline region are split into exclusive categories that are optimised for the best separation of the Higgs boson production processes. Two \ttH\ categories are defined to select events in which at least one top quark decays leptonically or in which both top quarks decay hadronically. The choice of background function and the estimation of the potential bias is carried out using data in dedicated control regions, where at least one of the two photons is required either to fail the tight identification (still passing loose identification) or isolation criteria. 
Figure~\ref{fig:Plots1} presents the invariant mass distributions in the two different event categories. The \ttHgg\ production cross section is measured to be: $ \sigma_{\ttH} \times \textrm{Br} (H \to \gamma \gamma) = -0.3^{+1.4}_{-1.1}$ fb, where the total uncertainty is dominated by statistics.

\section{\ttHleptons\ analysis overview}
The search for \ttH\ production using final states with multiple leptons, primarily targeting the decays $H \to WW$ and $H \to \tau\tau$, is performed with a dedicated cut-and-count analysis in four final states, categorised by the number and flavour of leptons: two same-charge light leptons with no hadronically-decaying $\tau$-lepton candidate ($2 \ell 0\tau_{\rm{had}}$), two same-charge light leptons with one hadronically-decaying $\tau$-lepton candidate ($2 \ell 1\tau_{\rm{had}}$), three light leptons ($3 \ell$), and four light leptons ($4 \ell$). 

\begin{figure*}[t]
\centering
\begin{overpic}[height=0.27\textwidth]{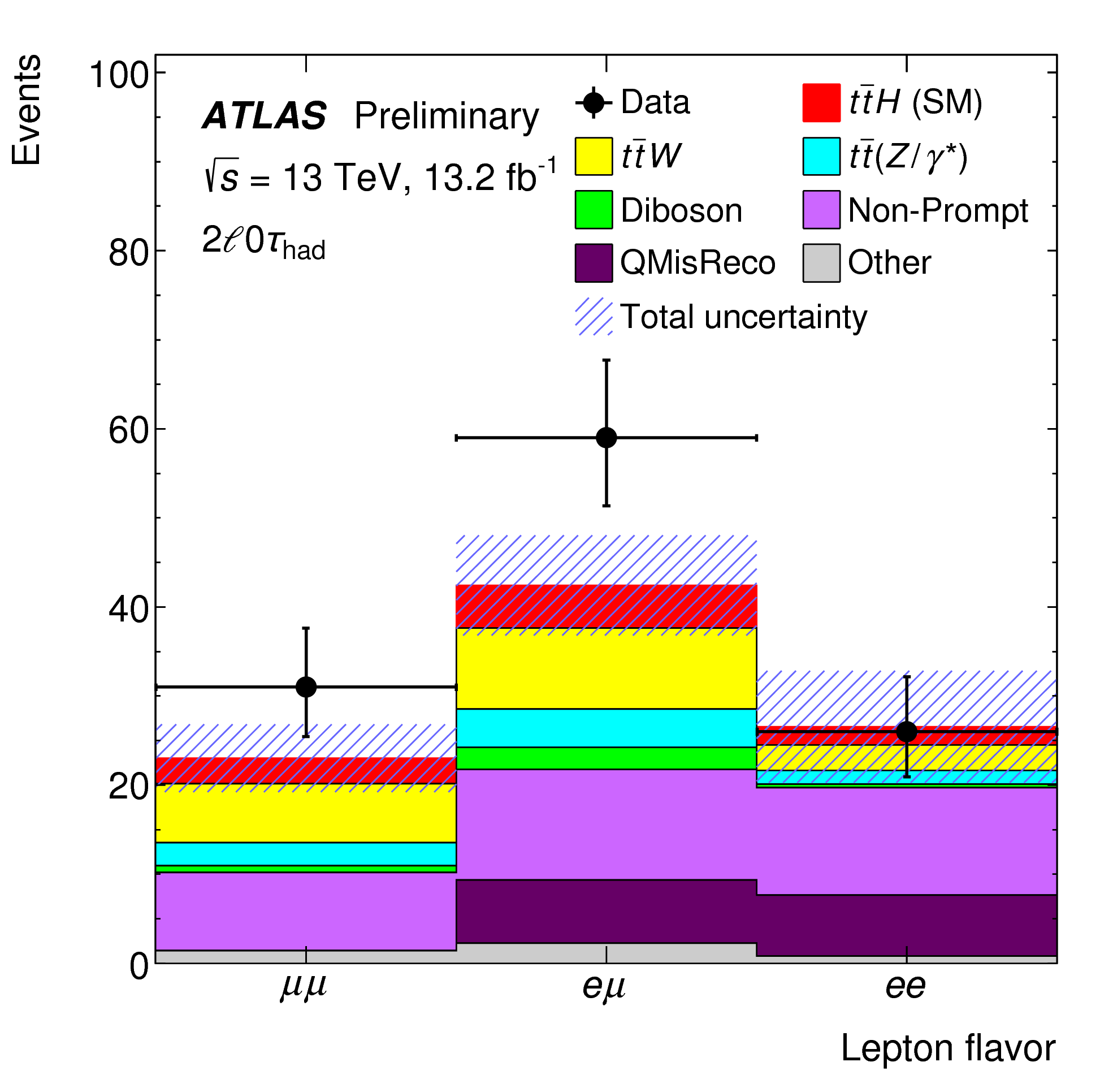}
\put(75,-7){(a)}
\end{overpic}
\qquad
\begin{overpic}[height=0.27\textwidth]{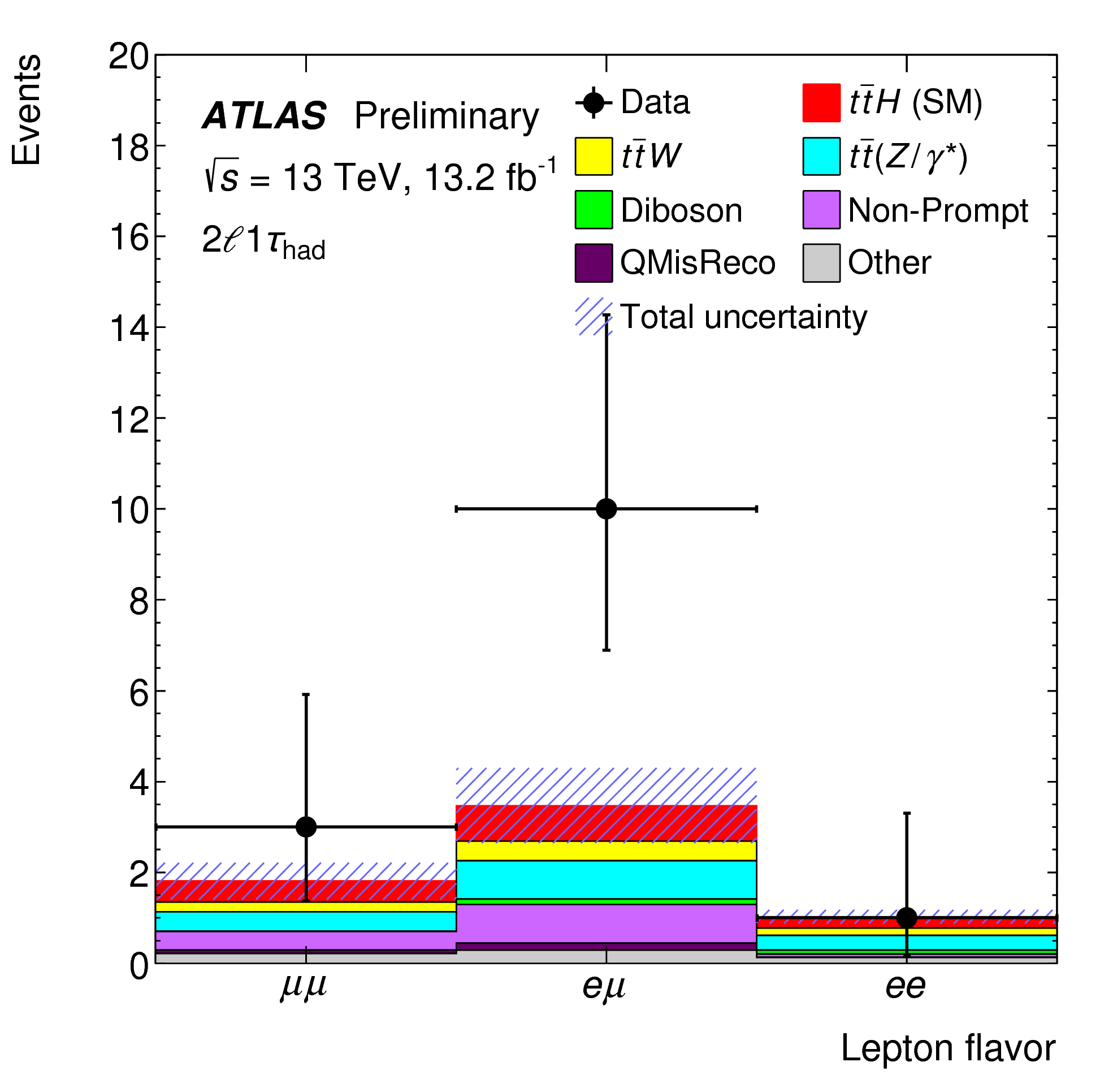}
\put(75,-7){(b)}
\end{overpic}
\qquad
\begin{overpic}[height=0.27\textwidth]{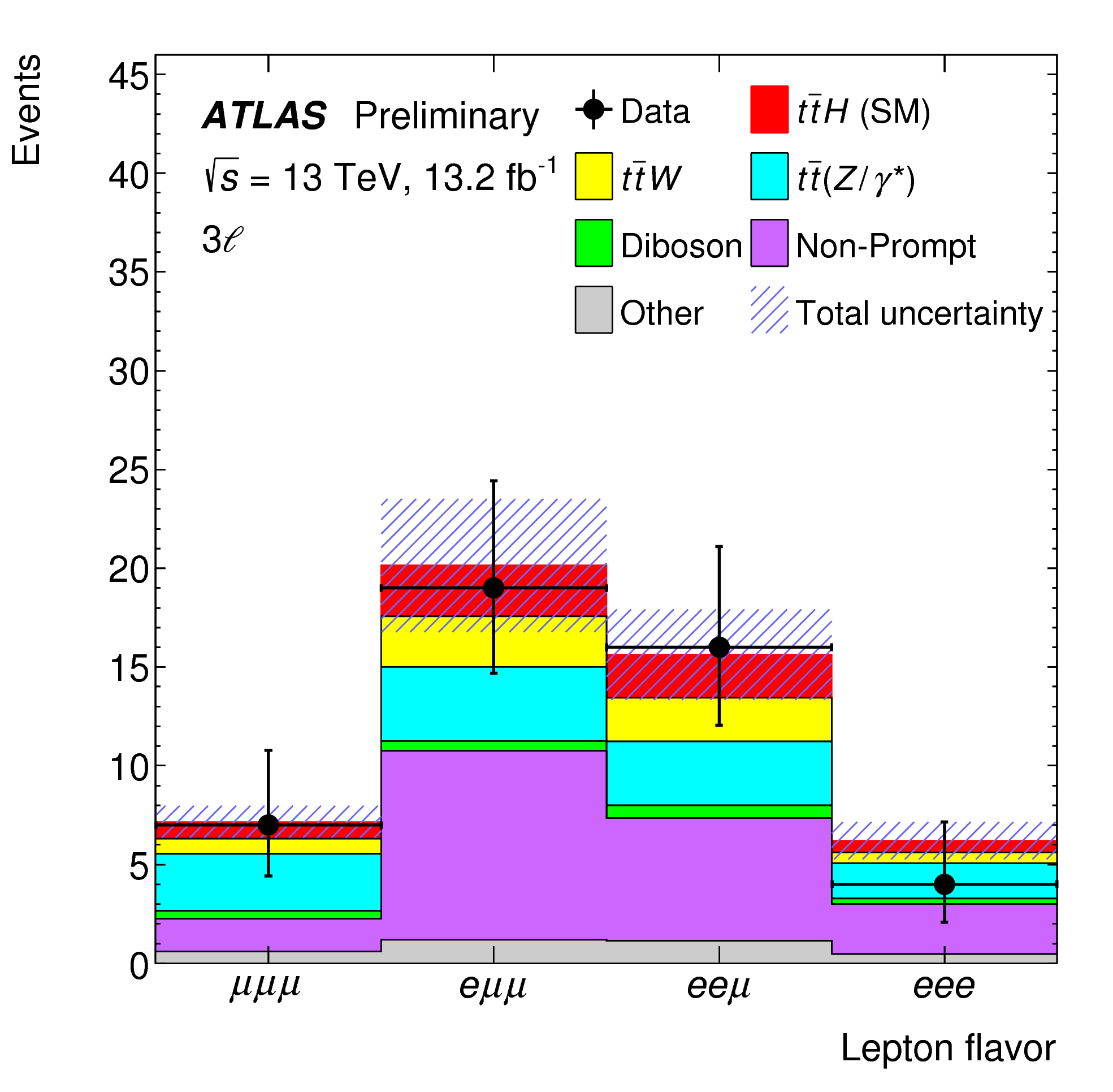}
\put(75,-7){(c)}
\end{overpic}
\caption{
Lepton flavour composition in the (a) $2 \ell 0\tau_{\rm{had}}$, (b) $2 \ell 1\tau_{\rm{had}}$ and (c) $3 \ell$ signal regions. The hatched region shows the total uncertainty on the background plus SM signal prediction in each bin~\cite{bib:ttH_multi}.
}
\label{fig:Plots2} 
\end{figure*}

The backgrounds for this search are categorised into those in which all the selected leptons are produced in decays of electroweak bosons or $\tau$-leptons (prompt leptons) and those in which at least one lepton arises from another source. In the latter case, the leptons arise from hadron decays or photon conversions (non-prompt), other interactions in detector material (charge mis-reconstruction or fake), or improper reconstruction of other particle species (fake). These backgrounds are estimated with a combination of simulation and data-driven techniques, and uncertainties in the non-prompt estimates have the largest effects on the background estimates. 
The best-fit value of the signal strength $\mu_{\ttH}$, combining all channels, is found to be $2.5 ^{+1.3}_{-1.1}$. Figure~\ref{fig:Plots2} shows the lepton flavour composition of the events in the $2 \ell 0\tau_{\rm{had}}$, $2 \ell 1\tau_{\rm{had}}$ and $3 \ell$ signal regions.

\section{\ttHbb\ analysis overview}
The \ttHbb\ search is designed for the $H \to b\bar{b}$ decay mode and uses events with at least one top quark decaying to an electron or muon. In order to take advantage of the higher jet and $b$-jet multiplicity of the \ttH\ signal process, the events are classified into exclusive regions based on the number of jets and the number of $b$-tagged jets. The regions where \ttH\ is enhanced relative to the backgrounds are referred to as signal regions, and a two-stage multivariate technique is used to separate the signal from the background, the latter being dominated by $t\bar{t}$+jets production. The remaining regions are taken as control regions, allowing a tighter constraint of backgrounds and systematic uncertainties in a combined fit with the signal regions. Dedicated systematic uncertainties affecting the modelling of the $t\bar{t}$+(heavy flavour) jets background are considered.

\begin{figure*}[hb]
\centering
\begin{overpic}[height=0.27\textwidth]{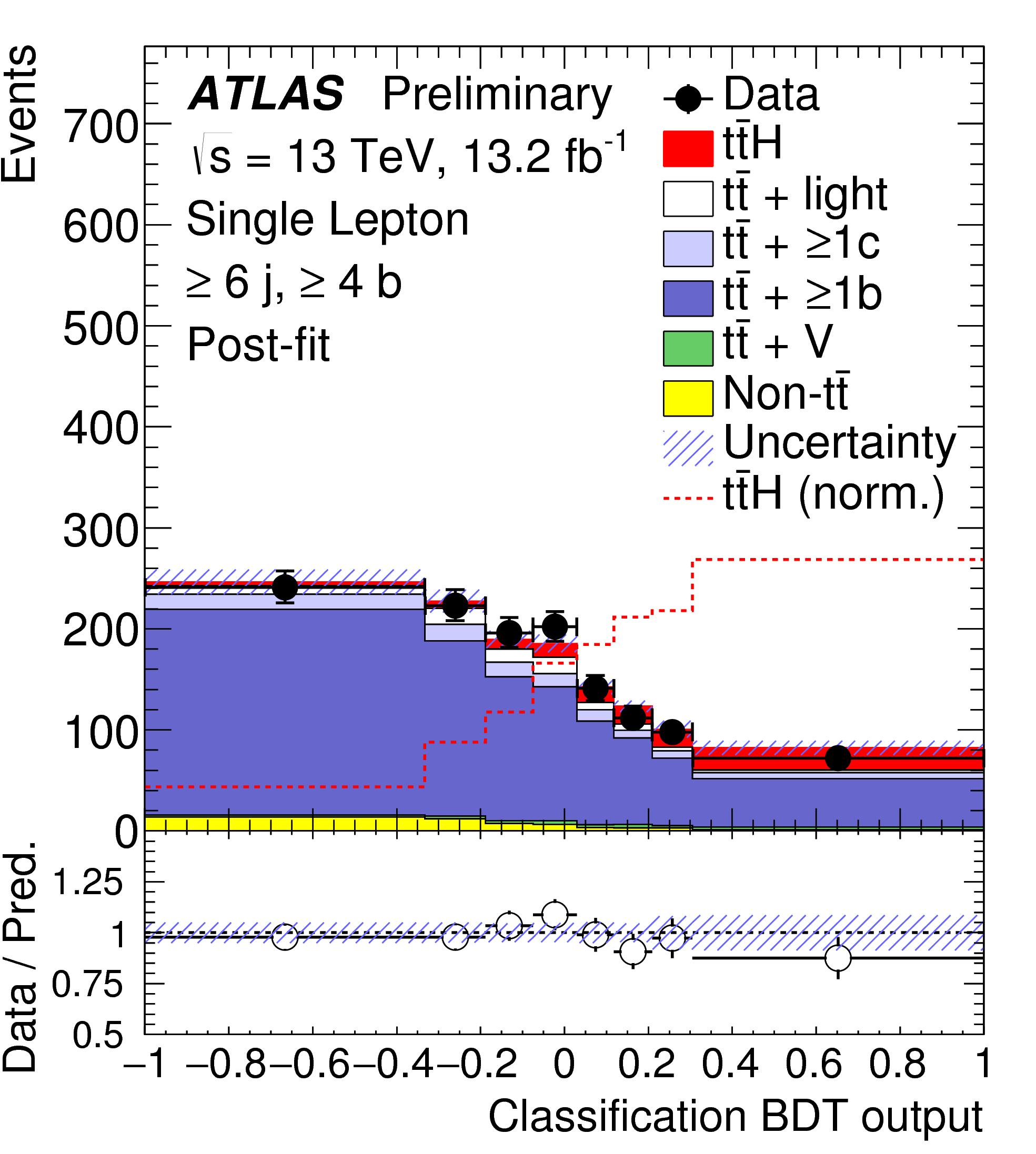}
\put(75,-7){(a)}
\end{overpic}
\qquad
\begin{overpic}[height=0.27\textwidth]{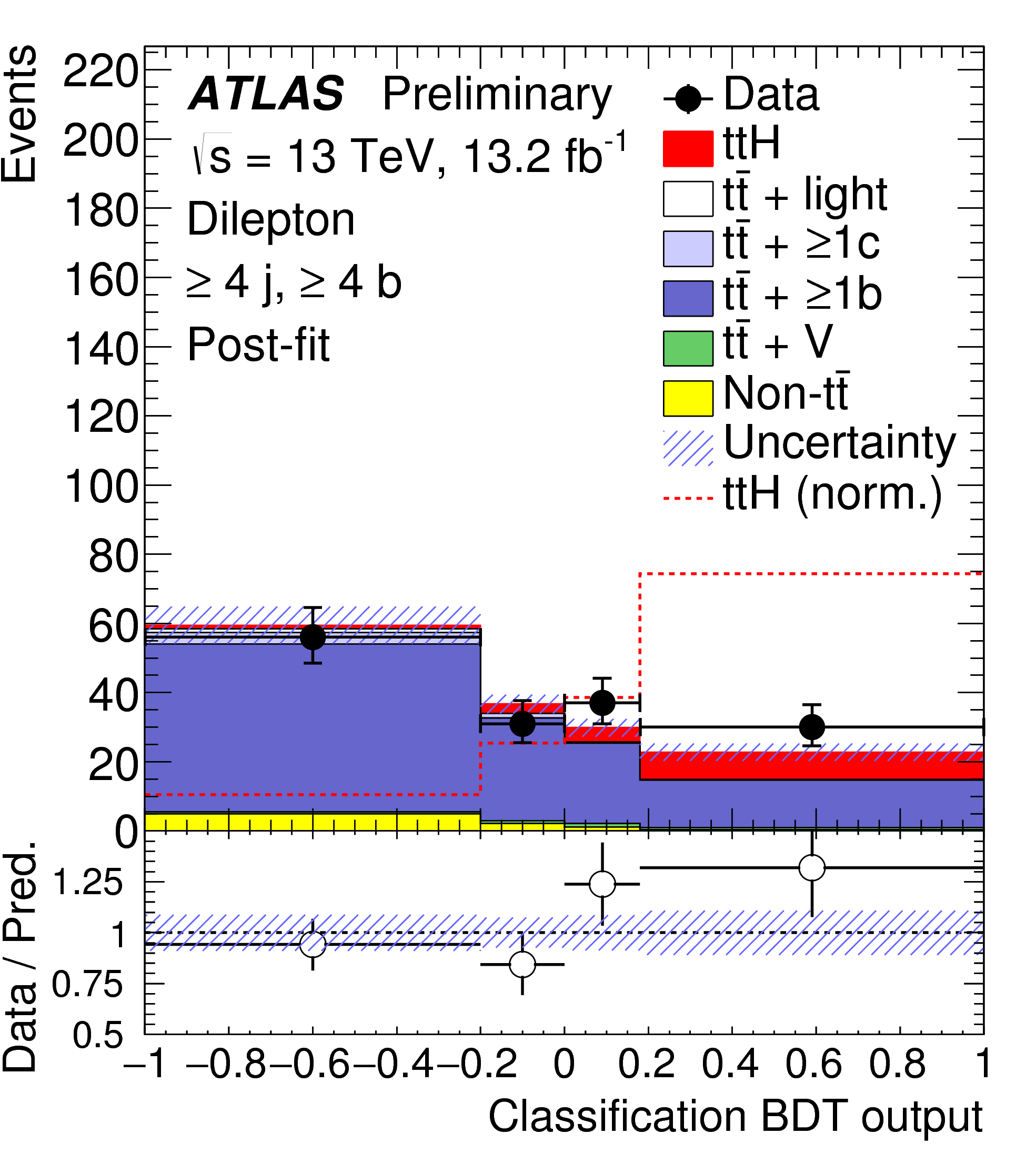}
\put(75,-7){(b)}
\end{overpic}
\qquad
\begin{overpic}[height=0.27\textwidth]{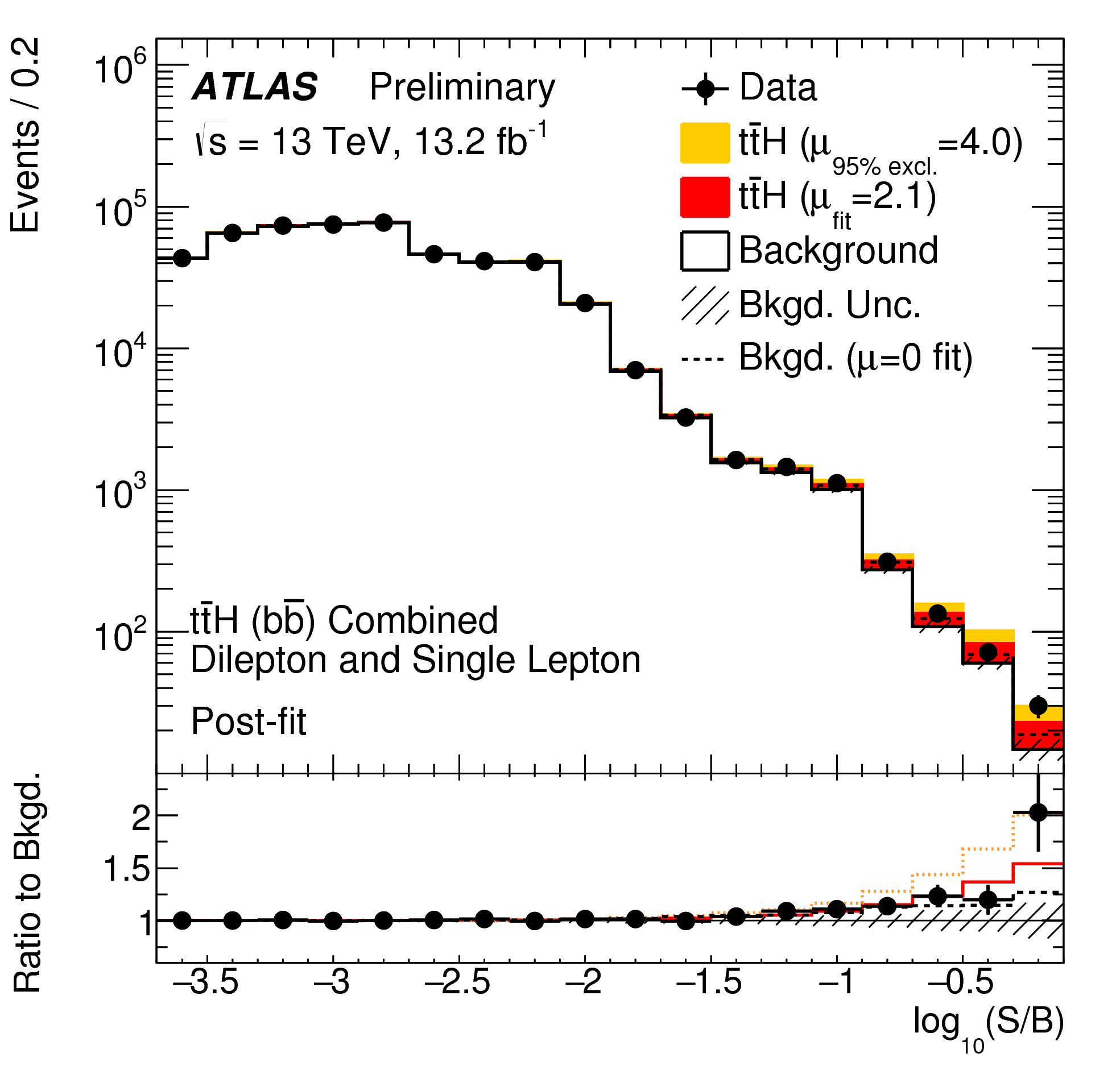}
\put(75,-7){(c)}
\end{overpic}
\caption{Comparison between data and prediction for the multivariate discriminant in the most signal-like region in the (a) single-lepton and (b) dilepton regions. (c) Post-fit yields of signal and total background per bin, ordered by log$(S/B)$, for all bins used in the combined fit of the single lepton and dilepton channels~\cite{bib:ttH_bb}.
}
\label{fig:Plots3} 
\end{figure*}

Figures~\ref{fig:Plots3}(a-b) show the distributions of the multivariate discriminant in the single-lepton and dilepton most signal-like regions, respectively. Figure~\ref{fig:Plots3}(c) shows the data compared to the post-fit prediction in each analysis bin considered, ordered by the $(S/B)$ ratio of the bins. The observed data are consistent with either the background-only hypothesis or with the the Standard Model \ttH\ prediction. The observed combined signal strength $\mu_{\ttH}$ is found to be $2.1 ^{+1.0}_{-0.9}$.

\section{Combination and conclusions}
\begin{figure*}[h]
\centering
\begin{overpic}[height=0.27\textwidth]{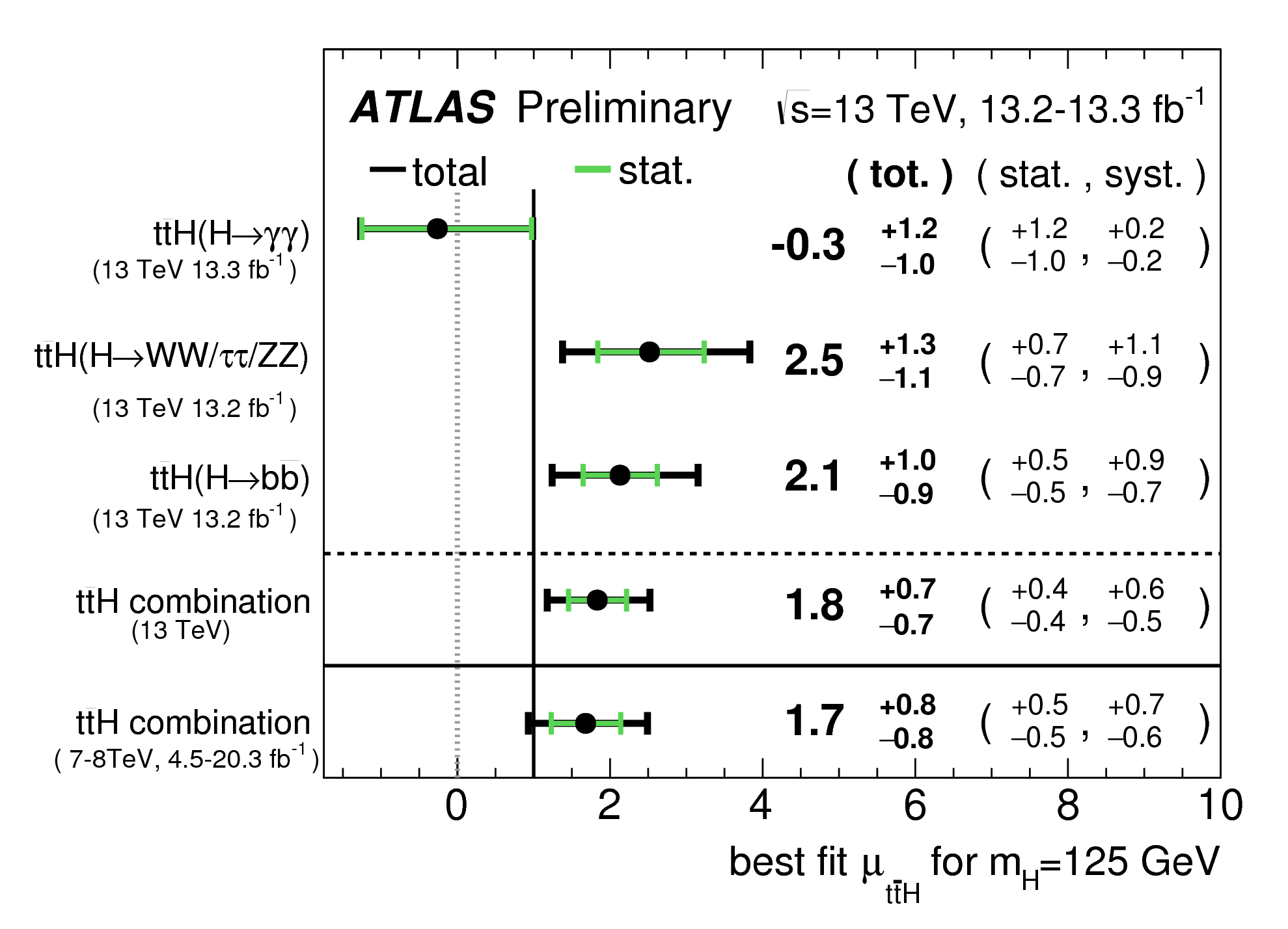}
\put(95,-7){(a)}
\end{overpic}
\qquad
\qquad
\begin{overpic}[height=0.27\textwidth]{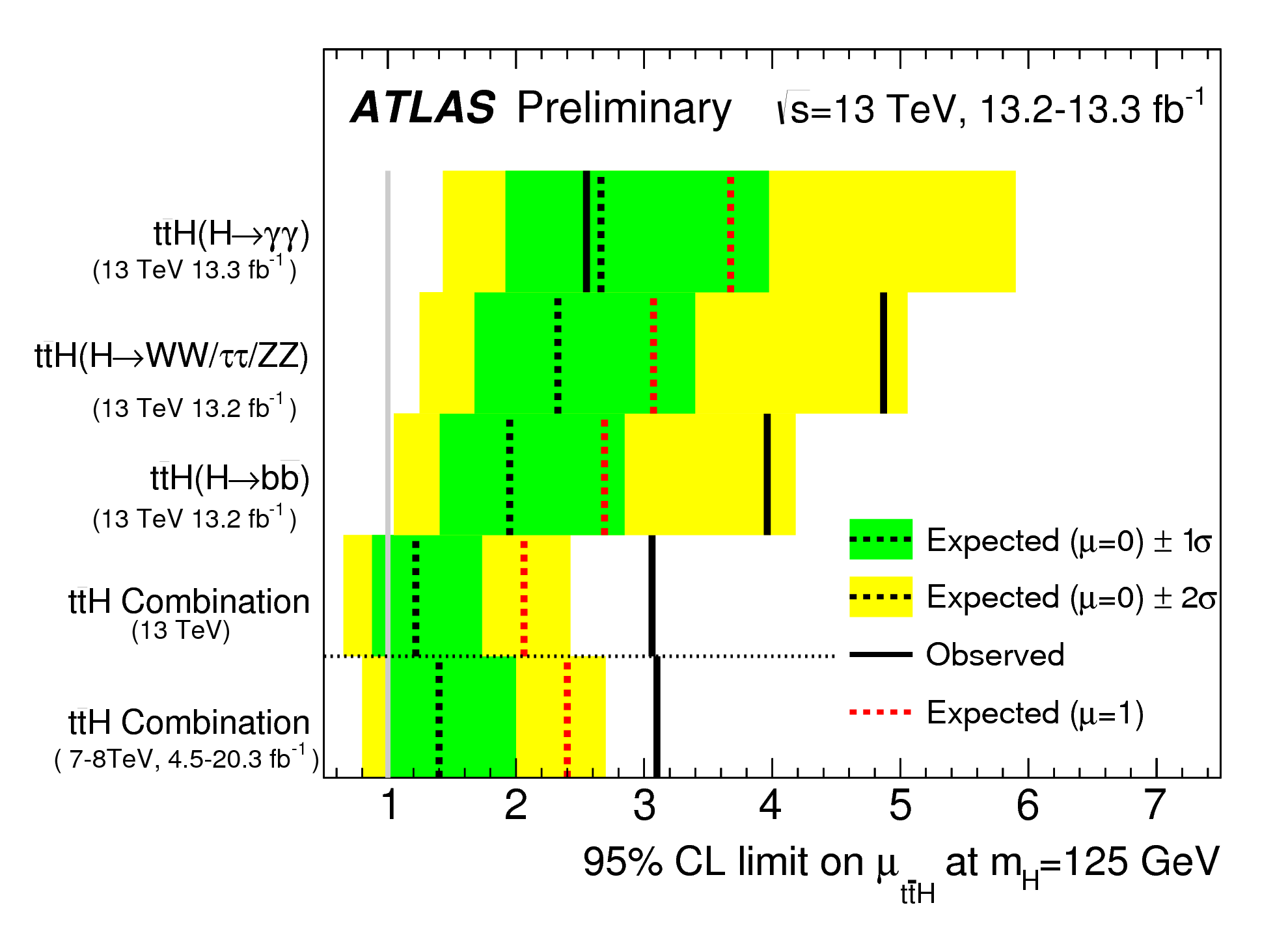}
\put(95,-7){(b)}
\end{overpic}
\caption{(a) Summary of the observed \ttH\ signal strength measurements from the individual analyses and for their combination, assuming $\mH = 125$~GeV. (b) Upper limits on the \ttH\ signal strength for the individual analyses as well as their combination at 95\% CL~\cite{bib:ttH_combo}.}
\label{fig:Plots4} 
\end{figure*}

The combination of the \ttH\ searches in the diphoton, multilepton, and \bbbar\ decay channels is performed using up to 13.3~fb$^{-1}$ of $pp$ collisions at $\sqrt{s}$ = 13 TeV collected with the ATLAS detector at the LHC. 
The combined \ttH\ signal strength ($\sigma / \sigma_{\rm{SM}}$) is found to be $1.8 \pm 0.7$, which corresponds to an observed significance of $2.8 \sigma$, where $1.8 \sigma$ would be expected in the presence of SM \ttH\ production. The summary of the observed \ttH\ signal strength is presented in Figure~\ref{fig:Plots4}(a). Upper limits on the \ttH\ signal strength for the individual analyses as well as their combination is presented in Figure~\ref{fig:Plots4}(b). All three analyses are within $1.5 \sigma$ of the central value, and the largest systematic uncertainty contribution is related to $t\bar{t}+b/c$ modelling uncertainties affecting the \ttHbb\ analysis. The sensitivity of the combination exceeds the Run 1 ATLAS expected significance of $1.5 \sigma$. 

Given the encouraging results obtained up to date by the ATLAS experiment in the searches for the SM \ttH\ production, the observation of \ttH\ production is expected at the end of the Run 2 at the LHC, and would be one of the main highlights of ATLAS physics program.

\end{document}

%% file: econfmacros.tex
\def\beq{\begin{equation}}
\def\eeq#1{\label{#1}\end{equation}}
\def\eeqn{\end{equation}}

\def\beqa{\begin{eqnarray}}
\def\eeqa#1{\label{#1}\end{eqnarray}}
\def\eeqan{\end{eqnarray}}

\let\bar=\overbar

\def\Dslash{\not{\hbox{\kern-4pt $D$}}}
\def\dslash{\not{\hbox{\kern-2pt $\del$}}}

\def\msb{{\bar{\ssstyle M \kern -1pt S}}}

\def\ttH{\ensuremath{t\bar{t}H}}

\def\mH{\ensuremath{m_{\rm H}}}
\def\bbbar{\ensuremath{b\bar{b}}}

\def\ttHbb{\ensuremath{t\bar{t}H(H \to b\bar{b})}}
\def\ttHgg{\ensuremath{t\bar{t}H(H \to \gamma \gamma ) }}
\def\ttHleptons{\ensuremath{t\bar{t}H(H \to WW, ZZ, \tau\tau ) }}

%% file: eprint.bbl
\begin{thebibliography}{99}


\bibitem{bib:atlas}
ATLAS Collaboration, JINST \textbf{3} S08003, (2008).

\bibitem{bib:ttH_gg}
ATLAS Collaboration, \textit{ATLAS-CONF-2016-067}, http://cds.cern.ch/record/2206210 (2016).

\bibitem{bib:ttH_multi}
ATLAS Collaboration, \textit{ATLAS-CONF-2016-058}, http://cds.cern.ch/record/2206153 (2016).

\bibitem{bib:ttH_bb}
ATLAS Collaboration, \textit{ATLAS-CONF-2016-080}, http://cds.cern.ch/record/2206255 (2016).

\bibitem{bib:ttH_combo}
ATLAS Collaboration, \textit{ATLAS-CONF-2016-068}, http://cds.cern.ch/record/2206211 (2016).


\end{thebibliography}
